# Singlet Deuteron, Dineutron and Neutral Nuclei.

S. B. Borzakov

FLNP JINR, Dubna, Russia

## Abstract

The existence of the dineutron was predicted over 70 years ago. At present, a number of experimental works confirm this assumption. By virtue of the principle of isotopic invariance, a singlet deuteron must also exist. The possibility of describing the neutron-proton interaction in the state at low energies as the excitation of a quasi-stationary level (singlet deuteron) lying below the deuteron decay threshold is discussed. The position, neutron and radiative widths of the level are determined by the scattering length, the effective radius, and the cross section for the radiative capture of neutrons by protons. Experiments to search for this level are discussed. The discovery of the singlet deuteron would be confirmation of the existence of the dineutron.

The idea of the existence of a dineutron was put forward as early as 1948 by N. Feather [1]. According to the Pauli principle, two neutrons can only be bound in the singlet ($^1S_0$) state. Feather estimated the dineutron lifetime (1-5 sec) and the maximum binding energy (3 MeV). An estimate of the binding energy was obtained from the condition for the β-decay of a dineutron into a deuteron.

To date, a number of experimental indications have been obtained in favor of the existence of the dineutron [2 - 6]. These works are not widely recognized because their results contradict the established paradigm. The point is that the traditional theory based on the model called "effective range theory" describes the low-energy scattering of two nucleons in the $^1S_0$ state as a manifestation of a virtual level with an energy of approximately -70 keV [7]. However, the physical meaning of the virtual level is unclear. In addition, this model does not allow for experimental verification. It must be understood that the theory of effective radius does not predict, but, strictly speaking, does not forbid the existence of the singlet deuteron. As noted in the book by Alfaro and Regge, it is impossible to determine the bound or virtual state describes the obtained experimental data on scattering [8].

The interaction of neutrons with protons in the $^1S_0$ state can be described using a dibaryon resonance with negative energy, that is, a quasi-stationary level located below the deuteron breakup threshold. The resonance energy is determined by the scattering length *a* and the effective radius $r_0$: $E_r = \dfrac{2}{a \cdot r_0}$. This formula was first obtained by S.T. Ma in 1953 by applying the R-matrix theory [9]. A similar approach is also used in [10]. In the case of interaction in the $^1S_0$ state, the scattering length is negative $a_s = -23.7$ fm, and, consequently, the resonance energy is also negative. In [11], this approach was applied to describe the interaction of neutrons with protons, taking into account two channels, scattering and radiative capture (see Appendix 1). Estimates give the following values of resonance characteristics:

$E_r$ = -1.3 MeV; $\Gamma_n$ (1 eV) = 10 keV; $\Gamma_\gamma$ = 20 eV.

At positive neutron energies, the description of the interaction with the help of resonance leads to the same results as the description with the help of the virtual level. But at energies below the deuteron photodisintegration threshold, there are significant differences. If this is a resonance (that is, a short-lived quasi-stationary state), then the following processes are possible: a) the emission of a cascade of gamma rays after the capture of neutrons by protons; b) resonant scattering of gamma quanta by the deuteron at an excitation energy $E^* = B_d + E_r < B_d$.

At present, the theory of strong interactions is based on quantum chromodynamics. As soon as ideas about quarks appeared, estimates of the masses of dibaryon states appeared. It was shown that the mass of the singlet dibaryon state is practically equal to the mass of the deuteron ($^3S_1$ - state) [12]. Recently, a number of works have appeared that estimate the contribution of six-quark states to the NN interaction and essentially predict the existence of a bound state of two nucleons in the $^1S_0$ state [13–16]. The theory of dibaryons is also being developed, in which the singlet virtual level is considered as a dibaryon resonance. They need experimental verification.

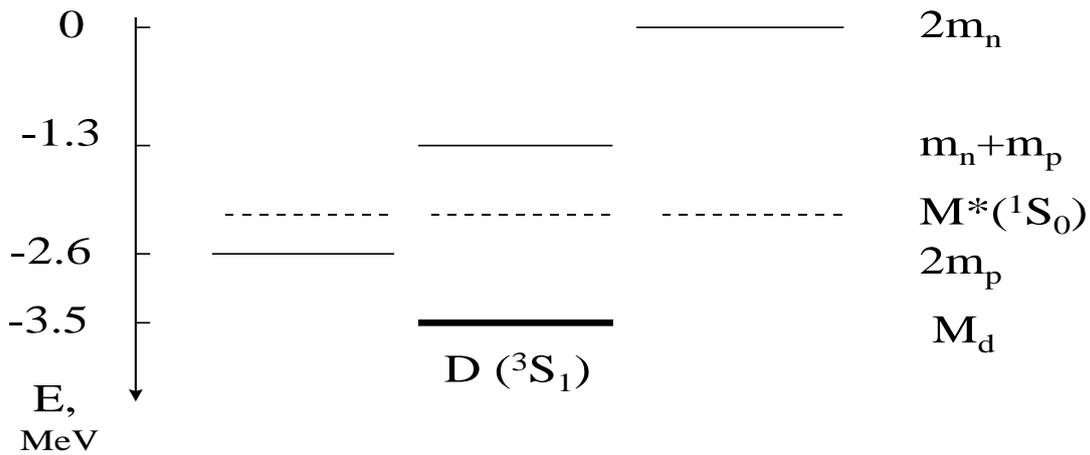

Fig. 1. The mass difference between different nucleon pairs.

The masses of different nucleon pairs are shown in Fig. 1. It is taken into account that the neutron mass exceeds the proton mass by 1.293 keV. The mass of the hypothetical dibaryon state is shown by the dotted line.

How to confirm or refute these ideas? A singlet deuteron can manifest itself in the radiative capture of neutrons by protons with the release of a cascade of two gamma quanta. One gamma quantum is emitted during the transition from the triplet state of the continuous spectrum to the singlet level, the second - during the transition from this state to the ground state (deuteron). The cross section of such a process is $10^3 - 10^4$ times smaller than the cross section with the emission of one gamma-quantum corresponding to the main transition [11, 15]. We have made attempts to find such a process. The registration of at least one gamma-quantum of the cascade would prove the existence of a singlet deuteron. Preliminary measurements were carried out at the IBR-2 reactor [17]. Measurements using HPGe, a high-resolution spectrometer with anti-Compton protection, were carried out at the Budapest Neutron Center [18]. A record statistics of gamma-quanta of the main transition with an energy of 2223.25 keV was collected - $2.8 \cdot 10^8$ counts. For the radiative capture cross section with the emission of a cascade of 2 photons, a limit of 2 μb was obtained in the narrow energy range of $E_\gamma$ = 2099–2209 keV, which corresponds to the binding energy range 15–125 keV [18]. The spectrum of gamma rays contains a line with an energy of 2212.9 keV, which can be interpreted as a manifestation of a cascade [19]. Unfortunately, the second line (with an energy of 10.4 keV) could not be detected in this experiment, since the detection threshold of the spectrometer was higher than this energy. Further research is required in a wider energy range with the organization of coincidences of two gamma quanta.

It was mentioned above that a number of experiments confirm the existence of the dineutron. But its characteristics remain unknown - the binding energy and lifetime. The dineutron can manifest itself in a reaction $^1_0n + ^2_1D \rightarrow ^2_0n + ^1_1p$. The original method to search for this reaction and determine the dineutron binding energy have been proposed in FLNP JINR. The idea is to detect a proton in a counter filled with deuterium [20]. Experiments with thermal neutrons did not lead to a positive result. Research with neutrons of higher energies is needed.

The next experiments are possible:

1. Search for a cascade of 2 gamma quanta with a total energy equal to the deuteron binding energy (2224 keV).

2. Search for resonant scattering of gamma rays on deuterons.

3. Search for a dineutron in a reaction $^1_0n + ^2_1D \rightarrow ^2_0n + ^1_1p$.

4. Search for inelastic scattering of neutrons by deuterons.

Thus, at present there are a number of works devoted to this topic and new experiments are planned. All these experiments do not require too expensive equipment and can be carried out in the near future. The list of references does not contain all the works related to this topic.

**Supplement 1.**

Derivation of a formula relating the scattering parameters (complex scattering length $A_s = a_s - i \cdot b$ and effective radius $r_0$) to the resonance characteristics.

$$F = \frac{1}{-\frac{1}{a-ib} + \frac{1}{2}r_0 k^2 - ik} = \frac{1}{2k}\frac{\frac{4}{r_0}k}{k^2 - \frac{2}{r_0 \cdot a} - i\frac{1}{2}(\frac{4k}{r_0} + \frac{4b}{r_0 \cdot a^2})}$$

$$= \frac{1}{2k}\frac{\Gamma_n}{E - E_r - i\frac{1}{2}(\Gamma_n + \Gamma_\gamma)}$$

$F$ – interaction amplitude; $\Gamma_n = \frac{4k}{\rho} \propto \sqrt{E_n}$ ; $\Gamma_\gamma = \frac{4b}{\rho a^2}$ ; $E_r = \frac{2}{ar}\frac{\hbar^2}{2\mu}$ ; $\frac{\hbar^2}{2\mu} = 41.47 MeV \cdot Fm^2$.

**Supplement 2.**

The problem of the dineutron's existence is connected with the question – do the neutral nuclei exist? There are number of interesting results in this direction have been obtained in the last days. 20 years ago, a wide resonance was caused by the work, in which 6 events were discovered, corresponding to the emission of a tetraneutron in the decay of [14]Be nuclei [1]. The formation of neutral nuclei was studied in [2–4]. The tetraneutron have been observed in [5]. It should be noted that, according to a number of theoretical works, the existence of neutral nuclei is impossible without the existence of a dineutron.

1. F.M. Marques et al., "Detection of Neutron Clasters", Phys. Rev. C65, 044006, 2002.
2. B.G. Novatski, E.Yu. Nikolski, S.B. Sakuta, D.I. Stepanov, "Possible observation of light neutral nuclei in the fission of [238]U by *α*-particles", JETPh Letters, v. 96, No. 5, p. 310-314, 2012.
3. B.G. Novatski, S.B. Sakuta, D.I. Stepanov, "The observation of light neutral nuclei in the fission of [238]U by *α*-particles with help of the [27]Al activation method", JETPh Letters, v. 98, No. 11, p. 747-751, 2013.
4. G.N. Dudkin, A.A. Garapatski, V.N. Padalko, "Arguments for Detecting Octaneutrons in Cluster Decay of [252]Cf Nuclei", e-arXiv, nucl-ex: 1306.4072, 2013.
5. T. Faestermann, A. Bergmaier, R. Gernhauser, D. Koll, M. Mahgoub, "Indications for a bound tetraneutron", Phys. Lett. B 824, 136799, 2022.